\documentclass[conference]{IEEEtran}
\usepackage{graphicx}
\usepackage{amsmath}
\usepackage{amssymb}
\usepackage{subfig}
\usepackage{wrapfig}

\usepackage[font=small,labelfont=bf]{caption}

\usepackage{array}
\newcolumntype{P}[1]{>{\centering\arraybackslash}p{#1}}
\newcolumntype{M}[1]{>{\centering\arraybackslash}m{#1}}

\newcommand{\ie}{{\em i.e.}}
\newcommand{\eg}{{\em e.g.}}

\DeclareMathOperator*{\argmin}{arg\,min}

\def\rrr#1\\{\par
\medskip\hbox{\vbox{\parindent=2em\hsize=6.12in
\hangindent=4em\hangafter=1#1}}}

\begin{document}

\title{Data Integrity Error Localization in Networked Systems with Missing Data 
\thanks{This work was supported by the
    US National Science Foundation under Grant OAC-1839900.}
}
\author{\IEEEauthorblockN{Yufeng Xin, Shih-Wen Fu, Anirban Mandal }
\IEEEauthorblockA{
RENCI, UNC - Chapel Hill\\
Chapel Hill, NC, USA\\
}
\and
\IEEEauthorblockN{Ryan Tanaka, Mats Rynge, Karan Vahi, Ewa Deelman}
\IEEEauthorblockA{
ISI, USC\\
Marina Del Rey, CA, USA\\
}
}

\maketitle
\thispagestyle{empty}

\begin{abstract}
Most recent network failure diagnosis systems focused on data center networks where 
complex measurement systems can be deployed to derive routing information and ensure network coverage 
in order to achieve accurate and fast fault localization. In this paper, we target wide-area networks that 
support data-intensive distributed applications. We first present a new multi-output prediction model 
that directly maps the application level observations to localize the system component failures. 
In reality, this application-centric approach may face the missing data challenge as some input (feature) data 
to the inference models may be missing due to incomplete or lost measurements in wide area networks. 
We show that the presented prediction model naturally allows the {\it multivariate} imputation to recover the missing data. 
We evaluate multiple imputation algorithms and show that the prediction performance can be improved significantly in 
a large-scale network. As far as we know, this is the first study on the missing data issue and applying imputation techniques in network failure localization.

\end{abstract}

\section{Introduction}
\label{sec:introduction}
Assurance of data integrity has been one of the most fundamental aspects of networked systems and Internet applications. 
Different mechanisms for error tolerance, detection, and mitigation have been widely implemented and deployed in different layers of 
the compute, storage, network, and applications systems. Unfortunately, these measures are not sufficient to address data corruptions 
in large-scale networks. For example, it is well known that the Ethernet CRC and TCP checksums are too small for 
modern data sizes~\cite{tcp:ccr2000}. Facebook recently reported a CPU bug that caused severe data corruptions in its hyper-scale data centers~\cite{facebook:cpu:2021}. 


Data integrity error is a representative ``gray'' network failure~\cite{GrayFailure:2017}, where 
network component faults are probabilistic and often evasive from the monitoring system. Gray failure diagnosis in large-scale networks 
is often a latent act after disastrous results from the applications and services. It remains a guessing art in most systems, which requires intensive 
manual debugging and a daunting amount of communication between operators 
from different organizations, which often takes days, primarily due to the infeasibility of accurate network models and 
incomplete coverage of system monitoring. 

In recent years, machine learning (ML) techniques have been applied with phenomenal success in addressing the failure diagnosis challenges.
The strength of ML models in learning the complex mapping between the failure root causes and the system level measurement 
observations effectively relieves the need for accurate domain models and full element level monitoring.  However, the majority of 
these systems targeted the data center networks, where detailed network topology and traffic routing information can be determined and 
expensive active and passive measurement systems can be deployed by the operators. The main 
technical contributions have been about developing scalable inference models and measurement systems of complete coverage for fine classification or regression performance~\cite{netbouncer:nsdi18,DeepView:NSDI18,arzani2018democratically}.

Accurate and fast fault localization requires complete network coverage from the measurement system within a limited time window. 
From the perspective of ML, training and test data sets with complete features are required for the model and inference. 
This is why substantial efforts were made in the existing work to develop complex measurement systems to ensure the network failure coverage. 

In this study, we target a completely different network environment, Internet scale network, where multiple networks spanning different administrative domains are used to support distributed applications. Different from the data center networks, while scalability might be more limited depending on topological and diagnostic granularity, challenges on obtaining available system information and measurement data are exacerbated by the multi-domain nature of the Internet and applications. In this ``opaque" network setting, deploying active probing to gather the network information and instrument the diagnosis in the production network is normally not feasible. It is also not realistic to deploy always-on passive monitoring system over the edges of the entire network.
 
On the other hand, modern Internet application software systems have built-in measurement and monitoring capabilities to ensure the application performance 
and mitigate the effects of failures in the underlying network system layer~\cite{IntegrityVerification:DataTransfer,swip:pearc:2019,iris:ictc21}. 
From the gray failure diagnosis perspective, the fundamental ML based solution approach appears to be a very viable choice with its promise in learning a model to map the observations to internal faulty behaviors of the network. In our case, the design goal is application-centric: to infer the fault information at the component level from the application level measurement and monitoring information. 

A big challenge in end host based application-centric solution approach is incomplete data, as discussed above. First of all, the end hosts deployed on the network 
by an application may not need all the network components in forwarding the traffic. Even if we limit our scope to exclude the part of the network not being covered,  
in both training data set and testing data set, some features' data may be missing in some data samples. In the extreme case, some features may be missed totally for the entire data set. 
This challenge of missing data is exacerbated when dealing with data integrity errors for which the failure rate is normally very low and the application traffic 
may be very imbalanced among the source and destination hosts at the edge of the network.
They may be caused by either incomplete coverage of the measurement, or the lost measurement records as the distributed measurement 
sub-system itself is not completely reliable in real-time, or there are just no measurements available for a part of application traffic during some time windows. 

Missing data has been a prominent research topic in statistics and is garnering more active research interests
in ML applications in the areas of medical science~\cite{DONDERS20061087} and sensor applications~\cite{missingdata:sensor:20}. The simple imputation techniques, like using the basic statistics (zero, mean, min, max, etc) of the existing feature data do not apply to our network failure diagnosis model because of the strong dependency between the application flows (features) and the system components (targets).   
The more suitable choices are the multivariate imputation algorithms that use the whole feature space to estimate (impute) the missing data in particular features.

In this paper, we first present a new multi-output ML prediction model that directly maps the application level observations to localize the system component failures. 
This model not only captures the fact that one faulty component would cause failures in multiple application flows, but also naturally allow the application of 
proven imputation methodologies to address the {\it missing data} challenge. Instead of pursuing more system information and complete measurement coverage, we focus on the multivariate imputation algorithms, parameter tuning, and quantifying their performance in improving the inference accuracy of failure localization. We also investigated the most recent algorithms based on the Generative Adversarial Nets (GAN) framework to generate the missing data~\cite{Yoon2018GAINMD,Awan2021ImputationOM}.

As far as we know, this is the first study on the missing data issue and applying imputation techniques in the area of network failure diagnosis. The evaluation results 
show satisfactory prediction accuracy. This model approach and missing data imputation results also present opportunities to the development and deployment of 
economical measurement capabilities in practical network settings.

\section{A Network Failure Localization Model}
\label{sec:fault}
\begin{figure}
  \begin{center}
    \includegraphics[width=0.48\textwidth]{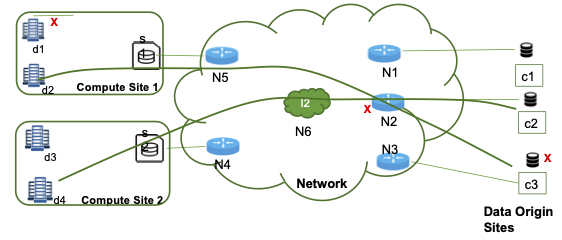}
  \end{center}
\caption{An Example Network.}
\label{fig:example}
\end{figure}

In this section, we present the inference model to localize the failures based on the path level measurements. 
Fig.~\ref{fig:example} shows a simple example to illustrate the problem. This networked system consists of several sites interconnected by a network cloud where only the network nodes and their interfaces are known (otherwise the failure localization is meaningless). 
A distributed application will incur traffic flows along an unknown path between pairs of end hosts. These flows are subject to the measurement of the applications to test if they are corrupted. Two such flows, $c3-d2$ and $c2-d4$ are shown in the figure. Intuitively, if both flows suffer from data corruption, the interface with the cross mark in Node $N2$ should be inferred to be the culprit.   

In most existing works on in gray failure localization, the network is modeled as a simple graph $G(V,E)$  
with a set of nodes $V$ connected by a set of links $E$. The bipartite mapping formula in Eq. (1) was used to capture the relationship between the link failures and the path level measurements~\cite{netbouncer:nsdi18,DeepView:NSDI18,arzani2018democratically}. 
The failure being considered is packet loss. 
The reasonings behind these models are similar: due to scalability or privacy constraints, 
monitoring every component of interest in a large-scale network is 
not feasible, while the path level measurement is more practical to deploy and instrument.  
\begin{flalign}\label{eq:prob}
\begin{aligned}
&Probability (No\ Failure\ in\ Path \ i) = \\
&\prod_{j \in Path\ i}Probability (component\ j\ is\ normal)
\end{aligned}
\end{flalign}
This can be transformed to a familiar linear regression model for every identifiable path in the network after taking log on the equations.
\begin{flalign}\label{eq:linear}
\begin{aligned}
&p_i = &\sum_{j \in Path\ i} c_j\ & \forall i \in P
\end{aligned}
\end{flalign}
Here $P$ represents the set of paths that are measured and a fundamental assumption underscoring these models is that routing of every path 
in $P$ over link set $E$ needs to be obtained. In addition, it was also assumed that the measurement system has the access to all network nodes to instrument path measurements, \ie, the source and destination of a path can be any nodes. In summary, to establish the regression model to obtain satisfactory inference performance, substantial efforts were made to (i) identify the routing of the paths, (ii) determine the path set for good coverage, and (iii) enable constant measurement of paths. Then the path measurements ($p_i$) will be obtained to estimate the link error probability ($c_j$) using this model.

In our wide-area multi-domain network setting, as we discussed in an earlier work~\cite{iris:ictc21}, it is not practical to identify the routing of the paths over the network and even the network topology beforehand. This means that it is difficult to establish a model like Eq. (\ref{eq:linear}). 
We also can not assume access to nodes in the network domains to instrument or measure paths.  

We hereafter distinguish between application end hosts (that generate and receive data) and networking devices (routers or switches), 
\ie, $V$ includes $H$ end hosts and $R$ routers. And we redefine $E$ to be the set of network components where failures are supposed to be 
localized, specifically all the network interfaces on $V$ and the end hosts $H$. We further constrain that only passive path measurements are available from certain applications on the end hosts, \ie, $P$ in our system only consists of paths originating from and ending in end hosts in $H$, which implies a much smaller identifiable path set. In the example network Fig.~\ref{fig:example}, none of the network nodes, $N1, \ldots, N6$, can be the source or destination of a path. And for a path between two end hosts, its routing is unknown. The only knowledge our model has is the bag of nodes and their interfaces for traffic forwarding.  

We further observe that one component failure (e.g., $x_j$) 
could cause multiple paths erroneous while one erroneous path may be the result of a failure at different components. The standard model Eq. (\ref{eq:linear}) ignores 
the correlation between multiple paths sharing a common component. We thereafter inverse the equations to represent the component failure probability as a function of 
the path failure probabilities as the following prediction model. 
 
\begin{flalign}\label{eq:inverse}
\begin{aligned}
Y = F(x_1, \cdots, x_p, \cdots, x_{|P|} ) \\
 = \sum_{p \in P} w_p x_p +w_0
\end{aligned}
\end{flalign}

Specifically, $Y$ represents a vector space $(y_1, \ldots, y_v, \ldots, y_{|V|})$ where $y_v$ represents the failure probability of component 
$v \in V$.  $X = (x_1, \ldots, x_p, \ldots, x_{|P|})$  forms the feature space that is defined by the combinations of the path failure probability. 
As shown in Eq. (\ref{eq:inverse}), we can further make it a linear regression model, which produces excellent performance as we will show in 
the evaluation section. We note, unlike in the existing work where failure localization is on network links, the network components in our model 
are the nodes and their interfaces because we assume the network topology is unknown.

Since any component failure only affects a small number of paths that go through it, plus multiple simultaneous failures are rare in reality, it is 
reasonable to expect that both the feature matrix and the coefficient matrix are sparse, representing the samples collected during one inference 
window. This suggests using the regularization technique to make most of the estimated coefficient to be zero. The most efficient technique 
to achieve this intention is to add a L1-norm constraint known as Lasso~\cite{DeepView:NSDI18}, where the regression optimization objective is defined as:    

\begin{flalign}\label{eq:lasso}
\begin{aligned}
\hat{W} =  \argmin_{W \in R^{|P|}}\vert\vert{\textbf{Y}-\textbf{X}W}\vert\vert _2^2 + \lambda \vert\vert{W}\vert\vert_1
\end{aligned}
\end{flalign}

Here $\textbf{Y}$ and $\textbf{X}$ are the sample matrices. This technique has proven extremely efficient in dealing with overfitting. 
The vector definition of $Y \in R^{|C|}$ means for each sample $n$, all entries but one in $Y^n$ are zeros. 
Compared to the scalar variable of a specific component failure probability, this multi-output model 
captures the independence between all the failures and would help the training and prediction quality.

\section{Missing Data and Imputation}
\label{sec:sl}
As discussed in Section~\ref{sec:introduction}, missing data is pervasive in reality due to lost or unavailable measurement data. 
It means that some samples, in the training set or the test set, have missing features. 
Using the example network in Fig.~\ref{fig:example} to illustrate, during a diagnosis time window, the application may not incur traffic between 
$c1$ and $d1$, or it never needs to transfer data between the origin sites, or the application measurement system may corrupt or lose some 
measurement data for some traffic flow. All these will lead to `holes' in the feature columns in the data sets. In the first two cases, 
entire feature columns will be missing in our model (Eq.~\ref{eq:inverse}).

Missing data can be categorized into three types: (i) the data is missing completely at random (MCAR), i.e. it does not depend on any of the observed and unobserved variables, (ii) the data is missing at random (MAR) if it is dependent only on the observed variables, (iii) the data is missing not at random (MNAR) if it is neither MCAR nor MAR, \ie, whether the data is missing depends on both observed variables and the unobserved variables. The majority of existing studies used the MCAR assumption~\cite{Yoon2018GAINMD}. 

There are many kinds of missing data recovery methods commonly used in the literature. These methods largely fall into two categories.

The {\it univariate} methods impute values in a  feature dimension using only non-missing values in that feature dimension. It simply replaces the missed 
values with certain statistics of the non-missing values such as the zero, mean, median, mode, max, or min. 

The {\it multivariate} imputation algorithms use the entire set of available feature dimensions to estimate the missing values based on the assumption of correlations between the feature dimensions. Each feature with missing values is modeled as a function of other features, and therefore the imputation itself is modeled as a regression problem that is trained and used to estimate the imputation. In order to achieve the best performance, especially to avoid the 
overfitting from certain features, it is conducted in a series of regression iterations: at each step, a feature is used as the output of other features and the resulting model is used to estimate the missing feature. After all features are processed or the designated max iteration is reached, the results of the final estimation are used to impute 
the missing data.

Our prediction model in Eq.~(\ref{eq:inverse})  uses the path (flow) measurements as the input. It naturally fits the multivariate imputation approach 
because the path failures caused by a common component failure are correlated. In contrast, the existing models based on Eq.~(\ref{eq:linear}) use the 
component failure as the input variables that are independent of each other, where the {\it multivariate} imputation does not make sense.
The {\it univariate} method is deemed not applicable due to the sparse nature of the feature matrix and lack of reasonable explanation.

Most recently,  the Generative Adversarial Nets {\it (GAN)} framework has shown good performance to generate the missing data.
In this model, the generator’s goal is to accurately impute missing data, and the discriminator’s goal is to distinguish between observed and imputed 
components. The discriminator is trained to minimize the classification loss (when classifying which components were observed and 
which have been imputed), and the generator is trained to maximize the discriminator’s misclassification rate. Thus, these 
two networks are trained using an adversarial process~\cite{Yoon2018GAINMD,Awan2021ImputationOM}.

There are off-the-shelf libraries that support both univariate and multivariate imputations in popular software packages like R and 
Scikit-learn~\cite{JSSv045i03,10.1371/journal.pone.0254720}. The imputation can also be performed multiple times with different 
random number seeds to generate multiple imputations. This is important if the statistical analysis is needed, \eg, in the medical domain. 

Most of existing missing data studies focus on minimizing the imputation errors of the data in the feature space. However the ultimate goal 
is the performance of the prediction models after missing data is imputed.

Corresponding to our model in Eq.~(\ref{eq:inverse}), missing data will cause values of some $x_p$ to be null. 
The feature space is defined in a $|P|$-dimensional space $\mathbf{X} = \mathbf{X_1} \times \ldots \times \mathbf{X_{|P|}}$. 
Following the MCAR assumption 
on the missing data, we can define a mask vector $M = (M_1, \ldots, M_{|P|})$ taking random values in ${(0, 1)}^{|P|}$.  
A sample vector $X = (X_1 \times \ldots \times X_{|P|})$ 
can be masked by $M$ to generate a corresponding sample vector with missing data $\tilde{X} = \tilde{X}_1 \times \ldots \times \tilde{X}_{|P|}$ as follows:

\[
\tilde{X_p} = 
\begin{cases}
  X_p & \text{if $M_p = 1$} \\
  null & \text{otherwise}
\end{cases}
\]

From an arbitrary missing rate $r \in (0, 1)$, a random mask vector $M_r$ can be created to emulate missing data from a given feature matrix $X$. 
For a particular missing feature $X_r \in X$, the imputation essentially creates a regression model that makes $X_r$ the output variable and all the 
other features the input variables.   

\begin{flalign}\label{eq:imputation}
\begin{aligned}
X_r = F(x_1, \cdots, x_p, \cdots, x_{|P|} ), \ p \neq r \\
\end{aligned}
\end{flalign}

At the end of the imputation, a recovered data set $\hat{X_r}$ is generated. The goal is to make these as close as possible.

Our main results are based on the MCAR missing data model, the {\it multivariate} imputation algorithms, and regularized regression model, 
which can be summarized in the following pipeline definition with Scikit\_Learn.
\begin{verbatim}
    estimator = make_pipeline(
        IterativeImputer(random_state=0, 
        		missing_values=np.nan, 
        		estimator=impute_estimator),
        PolynomialFeatures(poly),
        br_estimator
    )
\end{verbatim}

In the pipeline, the {\it impute\_estimator} specifies the regressor for missing data imputation and the {\it br\_estimator} specifies the regressor to infer the 
localized failure probability. We added a {PolynomialFeatures} element to evaluate if polynomials of higher degree perform better than the linear regressor.

This pipeline construct allows us to systematically evaluate the performance of multiple regressors in both {\it impute\_estimator} and {\it br\_estimator}, as well 
as tuning their hyperparameters. As we discussed earlier, in theory, Lasso should be a suitable regressor in both places. We also evaluated other popular 
regressors that include Ridge, BayesianRidge, ExtraTreesRegressor, and KNeighborsRegressor.

\section{Experiments and Evaluation}
\label{sec:evaluation}
\subsection{Emulation}
\begin{figure}[!ht]
\begin{center}
\includegraphics[width=0.36\textwidth]{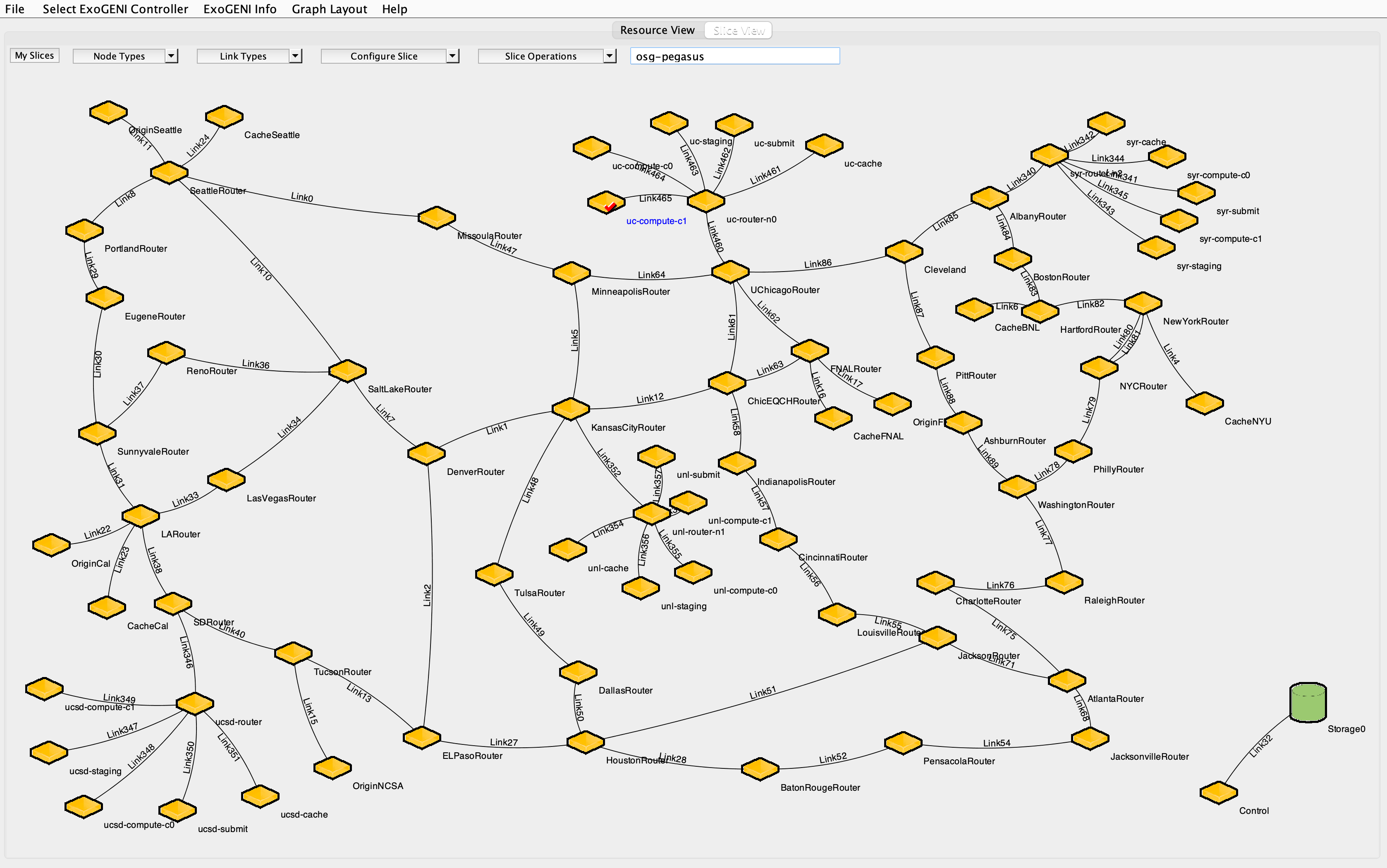}
\end{center}
\caption{Emulation Network Topology.}
\label{fig:topology}
\end{figure}

In this section, we validate the performance of missing data imputation with the proposed failure localization prediction 
model in an emulated network similar to the one shown in Fig.~\ref{fig:topology}. This topology mimics the Internet 2, the US Research 
and Education backbone network that has been used by many large-scale distributed middleware and applications.
It is created in a high-fidelity emulation environment we built in the NSF ExoGENI cloud testbed~\cite{iris:ictc21} 
that can automatically create a virtual network system with virtual machines (VMs) running full software stack, 
bootstrap the network routing, initiate data transfers, and inject arbitrary integrity errors into the virtual router interfaces 
and end hosts to generate labeled training data and test data.

The network consists of 15 end hosts that can originate and receive data transfers and 87 network interfaces that a simulated 
distributed application may transfer data over. The application can check the data integrity of every data transfer at the receiving end hosts.
During the emulation, errors with a given probability are injected into every end host and network interface in sequence, and one hundred 
data transfers are activated between all pairs of the end hosts in each round and the data transfer failure rates are computed. 
Therefore the training data set has 210 features, each of which represents a path between a pair of end hosts. We alter the error probability to generate 
multiple samples that are used to train the regression model.

To emulate the missing data, for a given missing rate, the mask vector $M$ is generated randomly and applied to the data site to create the 
missing data in the input matrix. The imputation performance is measured by the root mean square error (RMSE) between the original training data and the data recovered by the imputation algorithm. 


\subsection{Missing data imputation performance}
\label{subsec:im}
\begin{figure}[!ht]
    \subfloat[Imputation with Ridge\label{rmse:ridge}]{%
      \includegraphics[width=0.23\textwidth]{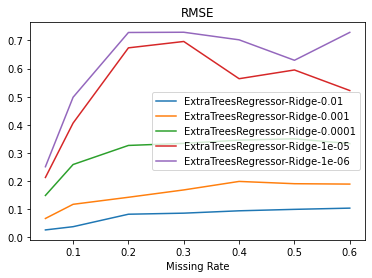}
    }
    \hfill
    \subfloat[Imputation with Lasso\label{rmse:lasso}]{%
      \includegraphics[width=0.23\textwidth]{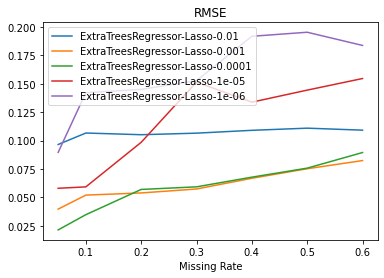}
    }
    \caption{Imputation with Regularized Regressors.}
    \label{fig:rmse:regu}
  \end{figure}

As we discussed in the last section, the regression models with a regularization term are normally suitable to sparse input matrices.
We evaluated two representative regularization methods, Ridge ($L_2$) and Lasso ($L_1$) with different penalty constant $\lambda$ in Fig.~\ref{fig:rmse:regu}.
When varying the missing rate from $5\%$ to $60\%$, $\lambda = 0.01$ for Ridge and $\lambda = 0.001$ and $\lambda = 0.0001$ for Lasso are the clear 
winners. Comparing the two figures, Lasso performs better than the Ridge counterpart.

\begin{figure}[!ht]
    \subfloat[Different Imputation Algorithms\label{rmse:im}]{%
      \includegraphics[width=0.23\textwidth]{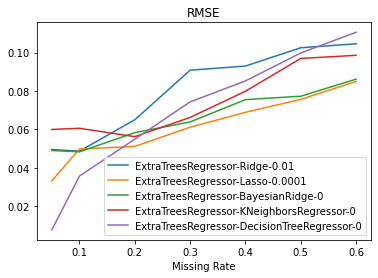}
    }
    \hfill
    \subfloat[Different Prediction Algorithms with Lasso Imputation\label{rmse:br}]{%
      \includegraphics[width=0.23\textwidth]{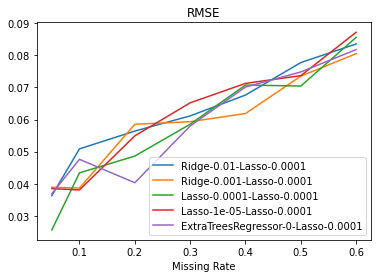}
    }
    \caption{Imputation Performance.}
    \label{fig:rmse}
  \end{figure}

We further evaluate three other regressors against the tuned Lasso and Ridge regressors in Fig.~\ref{rmse:im}. It confirms that the Lasso with a small 
penalty constant ($\lambda = 0.0001$) outperforms all other regressors except for the cases of very low missing rate. The BayesianRidge regressor performs 
closest to Lasso, the K-Neighbors regressor the second, and the ExtraTrees regressor the next. When the missing rate is low ($<20\%$), the ExtraTrees regressor 
performs much better than the other algorithms.

Since we implement the imputation regressor in the pipeline with the ultimate prediction model, we also tried different prediction algorithms 
with the best Lasso imputation regressor. From Fig.~\ref{rmse:br}, their impacts on the imputation performance are small, though different 
with different missing rates.
 
As we discussed in the last section, the latest GAN based imputation algorithms showed some promising results in the medical domain. 
We applied the Generative Adversarial Imputation Nets (GAIN) developed in~\cite{Yoon2018GAINMD} to our data set and compared its performance 
with Lasso in Table~\ref{tab:gain}. The results actually do not show that the GAIN model outperforms the Lasso algorithm. We will leave the further validation and 
customization of GAN framework to our future work.
  
 \begin{table}[!ht]
\caption{RMSE vs. Missing Rate}
\label{tab:gain}
\begin{center}
\begin{tabular}{ |c|c|c|c|c|c|c| } 
 \hline
  & 0.1 & 0.2 & 0.3 & 0.4 & 0.5 & 0.6\\ 
 \hline
  \hline
 GAIN & 0.1872 & 0.1794 & 0.214 & 0.2194 & 0.2666 & 0.2986 \\ 
 \hline
 Lasso & 0.05 & 0.0512 & 0.0612 & 0.0689 &0.0756 & 0.085\\
 \hline
\end{tabular}
\end{center}
\end{table} 

\subsection{Inference performance with imputed missing data}
\begin{figure}[!ht]
    \subfloat[Prediction Score\label{score:t}]{%
      \includegraphics[width=0.23\textwidth]{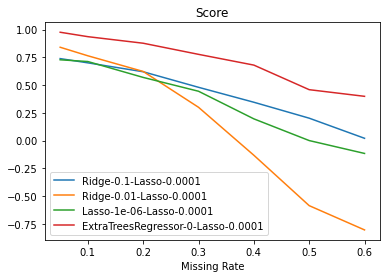}
    }
    \hfill
    \subfloat[Prediction MSE\label{mse:t}]{%
      \includegraphics[width=0.23\textwidth]{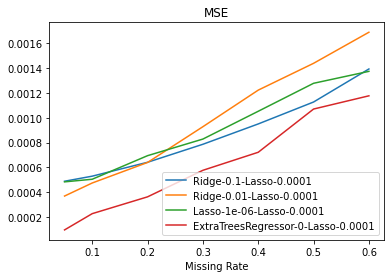}
    }
    \caption{Prediction Performance with Missing Data Imputation.}
    \label{fig:score}
  \end{figure}

Many missing data imputation studies stopped after evaluating imputation performance as we did in subsection~\ref{subsec:im}. 
However, our ultimate goal is to improve the inference accuracy of failure localization. In Fig.~\ref{fig:score}, we first evaluate the 
regression performance with different prediction algorithms on the best Lasso imputation regressor in terms of the 
coefficient of determination $R^2$ (Score) and the mean squared error (MSE) on the predicted output, the component 
failure probability vector $Y$ in Eq.~(\ref{eq:inverse}). We recall that higher score (best 1) and lower MSE mean better regression performance. 

The two metrics show that the ExtraTrees regressor actually achieves the highest score and lowest MSE. The Ridge regressor with a small penalty constant ($0.01$) results in the worst performance. The Ridge with a bigger penalty constant ($0.1$) and Lasso with a very small penalty constant ($0.000001$) result in a similar performance, not far from the ExtraTrees algorithm. 
  
\begin{figure}[!ht]
    \subfloat[Top-1 Accuracy\label{a:t:1}]{%
      \includegraphics[width=0.23\textwidth]{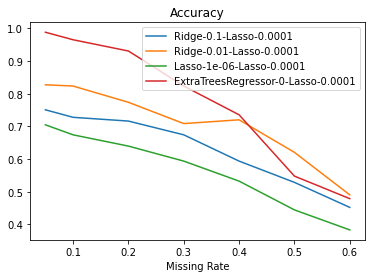}
    }
    \hfill
     \subfloat[Top-2 Accuracy\label{a:t:2}]{%
      \includegraphics[width=0.23\textwidth]{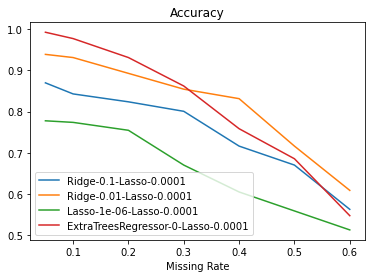}
    }
    \hfill
      \subfloat[Top-3 Accuracy\label{a:t:3}]{%
      \includegraphics[width=0.23\textwidth]{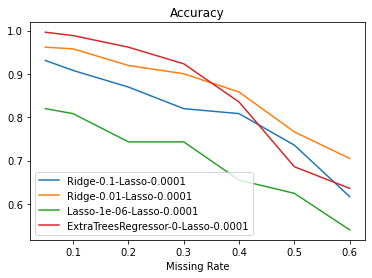}
    }
    \hfill
   \subfloat[Top-4 Accuracy\label{a:t:4}]{%
      \includegraphics[width=0.23\textwidth]{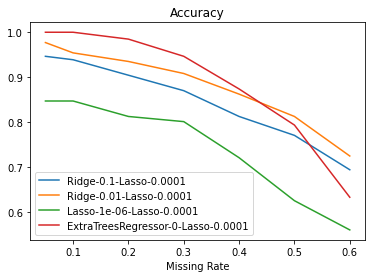}
    }
    
    \caption{Top-k Localization Accuracy with Imputed Missing Data.}
    \label{fig:topk}
  \end{figure}

We proceed to attempt to determine the failure locations with the above combinations of prediction and imputation algorithms. For each test sample, after the predicted component failure probability vector $\hat{Y}$ is calculated, we sort the vector elements in descending order. In anticipating that higher missing rate will likely significantly deteriorate the localization accuracy, we evaluate the $Top-K$ accuracy with small $k\le4$, which counts the correct label falls in the first $k$ elements in the sorted $Y$ vector. Fig.~(\ref{fig:topk}) shows the localization accuracy when $k=1,2,3,4$. 
  
The results are promising. For the exact match ($Top-1$) in Fig.~(\ref{a:t:1}), two best algorithms achieves over $50\%$ accuracy at the missing rate $60\%$. When the missing rate is higher than $40\%$, the Ridge (0.1)+Lasso combination starts to outperform the ExtraTrees+Lasso algorithm, whose performance drops below the Ridge (0.01)+Lasso algorithm when $k>1$. The results show some disparities from the regression performance results in Fig.~\ref{fig:score}. 
   
\section{Conclusions and Future Work}
\label{sec:future}
In this paper, we studied the diagnosis of data integrity errors in a wide-area network that supports data-intensive distributed applications. 
Due to its multi-domain nature, information of the network topology and traffic routing, and network layer measurements are either not available or are too expensive to obtain. Nevertheless, it is viable to construct Machine Learning models that rely on the application layer 
measurements to infer the failures inside the network. We first present a new multi-output ML prediction model that directly maps the application level measurements to the possible failure locations at the network components. In reality, this application-centric approach may face the {\it missing data} challenge as some input (feature) data to the inference models may be missing due to incomplete or lost measurements in the wide-area networks. Missing data and the associated imputation techniques have been prominent research topics in statistics and are garnering more active research interests in ML applications as it is a pervasive problem in reality.
 
As our prediction model uses path measurements as input, it allows the multivariate imputation because path failures caused by a common component failure are correlated. We introduced several imputation algorithms under different missing data scenarios. Using a high-fidelity emulation environment we built using a Cloud testbed, we evaluated the performance of the prediction model and 
the imputation techniques. The results showed fine-tuned regression model with regularization is very efficient in terms of missing data recovery performance and failure localization prediction accuracy.   

For our future work, we plan to experiment with larger networks having different graph properties and GAN based imputation methods. We note full network coverage is not the focus of this study. As traffic in a particular application may not pass through all the network components, we will explore mechanisms to federate measurements from multiple applications to achieve higher network failure diagnosis coverage.

\section*{Acknowledgments}
This work is funded by NSF award OAC-1839900. Results were obtained using the ExoGENI testbed supported by NSF.

\bibliographystyle{IEEEtran}
\bibliography{iris_sl,iris}


\end{document}